# ILCWS08 Test Beam Summary

Erik Ramberg

Fermi National Accelerator Laboratory
Batavia, Illinois USA

A summary is given of the high energy test beam facilities around the world. Attention is placed on the capabilities and availability of each. A short description is given of what kind of additional facilities are required in the future to support ILC detector research.

## 1  Motivation

Facilities with active beam for testing particle detectors play a sometimes subtle, but always crucial, link between conception and implementation of a particle physics experiment. Besides being the proving ground where a decision on detector technology is made, they also typically play a dominant role in calibrating freshly built detectors and sometimes in resolving systematic unknowns about detector performance under unusual conditions. In addition to these technical roles, any test beam manager can tell you about their important role in motivating a detector collaboration to meet realistic goals and in educating students new to the field.

The ILC detector R&D groups will certainly follow these paths, given that they propose to significantly push beyond the current limitations of both tracking and calorimetry. This summary outlines the status of test beam facilities in the three regions of Europe, Asia and North America, and gives a short outlook on what new facilities are needed.

## 2  Facilities Worldwide

The LCWS08 workshop heard short summaries of the status of 9 test beam facilities around the world. Table 1 shows some operating characteristics of these facilities, as well as two others.

### 2.1  European test beam facilities (Presentation by Lucie Linssen of CERN)[1]

Europe has two main centers for high energy test beams – CERN and DESY, which complement each other nicely. CERN has a large number of active beamlines and can reach the highest energies in the world. DESY has a very clean lower energy dedicated electron beam, with a pixel telescope and superconducting solenoid for supporting advanced detector R&D.

CERN PS area: There are two distinct areas at CERN where test beams are located. The first is in the East area, with beams delivered from the 24 GeV Proton Synchrotron (PS). There are 4 multi-purpose secondary beamlines: T7 (<10 GeV), T9 (<15 GeV), T10 (<7 GeV) and T11 (<3.5 GeV). There is one dedicated primary beamline (T8) for the DIRAC experiment.  In addition, the PS East area contains an irradiation facility in the T7 beamline that can deliver 1-10 x $10^{13}$ protons/(cm$^2$-hr) or 3-10 x $10^{11}$ neutrons/(cm$^2$-hr).  The user installation areas for these beamlines have beam profile measurements and threshold Cerenkov counter capability for particle i.d.  They cannot support infrastructure for large detectors or cryogenic installations.

## BEAM TEST FACILITIES

| Laboratory | Energy Range | Particles | Availability |
|---|---|---|---|
| CERN PS | 1 - 15 GeV | e, h, m | LHC absolute priority |
| CERN SPS | 10 - 400 GeV | e, h, m | LHC absolute priority |
| DESY | 1 - 6.5 GeV | e$^-$ | ~10 month per year |
| Fermilab | 1-120 | e, p, K, p; m | continuous (5%), except summer shutdown |
| Frascati | 25-750 MeV | e | 6 months per year |
| IHEP Beijing | 1.1-1.5 GeV (primary) 0.4-1.2 GeV (secondary) | e$^\pm$, e$^\pm$, p, p | Continuous after March 2008 |
| IHEP Protvino | 1-45 GeV | e, p, K, p; m | one month, twice per year |
| J-PARC | | | Available in 2009 |
| KEK Fuji | 0.5 - 3.4 GeV | e | Available fall 2007, 240 days/year |
| LBNL | 1.5 GeV < 55 MeV < 30 MeV | e p n | Continuous |
| SLAC | 28.5 GeV (primary) 1.0 - 20 GeV (secondary) | e, e$^\pm$, p$^\pm$, p | Parasitic to Pep II, non-concurrent with LCLS |

Table 1.  Description of world-wide test beam facilities (from I. Gregor talk)

CERN SPS area:  In the north, the Super Proton Synchrotron (SPS) delivers 450 GeV beam onto targets, from which 4 secondary beams are steered to test beam user areas in the EHN1 building.  The duty cycle for this beam is a 4.8 second flattop every 16.8 seconds.  This area has seen extensive use in recent years, with a wide variety of beam conditions, with energies typically in the 10-400 GeV range and sometimes very high rates.  The resolution for secondary beams is typically 1.5%.   Particle separation or identification has been supported wherever possible.  Support for installations includes counting houses and racks, beam profile measurements, scanning tables, threshold Cerenkov counters, CEDAR counters (differential Cerenkov), spectrometers and electromagnetic calorimeters.  Magnets and cryogenic installations are supported as well and there is a lot of area to support large detectors. In 2008 there were 32 SPS user groups, of which about 50% were LHC oriented. Both vertex, tracking and calorimetry tests for ILC detectors were supported this last year.

DESY: The test beamlines at DESY originate from the DESY II electron beam hitting a thin (10 μm) carbon fiber situated in the vacuum beamline. This creates bremsstrahlung photons which are subsequently converted by a variety of thin user-choice targets. The tertiary electron beams of 1-6 GeV energy are steered and collimated into three user areas: T21, T22 and T24. The beam arrives every fourth PETRA cycle, or 320 msec. Besides standard user infrastructure such as gas lines, translation stages and magnet controls, DESY hosts two very significant detector testing structures funded by the EUDET establishment. The first is a MAPS pixel telescope, which has been tested both at DESY and CERN. The second is a 1 Tesla, 85 cm diameter superconducting solenoid, for silicon vertex and TPC studies. The Linear Collider TPC collaboration plans to bring their large prototype detector to DESY for tests, with the goal of demonstrating proof of ILC scale momentum resolution in a full solenoid volume.

## 2.2   Asian test beam facilities (Presentation by Osamu Tajima of KEK)[2]

It is a time for transition in Asian beam test facilities as KEK ramps up their electron beam and operations are foreseen at J-PARC. The IHEP-Beijing facility shuts down to update their accelerator. And IHEP-Protvino focuses on tests for the new FAIR laboratory.

KEK, Japan: The KEK test beam facility celebrated its first year of operation in October. In that year, they performed 15 beam tests – a record for a single beamline. Bremsstrahlung photons are generated from the KEKB beamline and converted into 3.5 GeV electrons, which are steered through a 'roller-coaster' set of magnets into a small user area. The theme for creating this facility was 'recycle'. The beam rate is 20 Hz, with a 3 cm profile and momentum spread of 1%. Some notable detectors tested include:
- Silicon vertex detector for sBelle
- An aerogel RICH, read out by Hybrid Avalanche Photodiode technology
- Beam profiles measured in the ATLAS SCT silicon tracker
- Measurements on an ILC electromagnetic calorimeter

J-PARC, Japan: A significant new addition to the capabilities in Japan is the test beam area being built in the Hadron Hall at J-PARC. The primary proton beam of 30 GeV interacts in a beryllium target, from which secondaries are extracted at 50 degrees. A yield calculation indicates about 10 pions per 100 MeV interval in the 1-3 GeV range. Operations are expected to begin in 2010.

IHEP Beijing, China: The Beijing Test Beam Facility (BTF) season ended in March, 2008, with notable tests of the low energy X-ray telescope for HXMT, tests of the multi-gap RPC for STAR and a test of a CVD diamond film detector for BEPCIII. A long shutdown for upgrades is expected to be complete in 2010. The 1-1.5 GeV electron and secondary beam will have improved optics, beam monitors and alignment scheme. A TPC and GEM detector will be installed to get improved particle identification and momentum measurement resolution of 0.5%.

IHEP Protvino, Russia: At least four high and low intensity beamlines are available at Protvino. They deliver the widest range of secondaries in Asia. Beams include 2-19 GeV electrons, up to 34 GeV pions and up to 50 GeV protons. Rates can be from 100 to 1000 Hz, with a cycle time of 1.8 seconds every 10 seconds. Beam monitoring includes time-of-flight, threshold and differential Cerenkov systems, MWPC's and hodoscocpes. Beam time is available in two 1 month periods in April and then November. Beamline 2 will be intensively used for tests of the PANDA experiment of the FAIR facility in Darmstadt. Detectors to be tested include lead tungstate crystals, shashlyk calorimetry and a forward microstrip vertex detector.

## 2.3   North American test beam facilities (presentation by Doug Jensen of FNAL)[3]

In North America, only Fermilab is currently operating a general purpose high energy test beam facility. A proposal has been made to create a new facility at SLAC.

SLAC: At SLAC a new era has begun, with the planned startup of the Linear Coherent Light Source to take place this year. The LCLS takes over the last half of SLAC's famous linear accelerator and, without modification, would prohibit extraction of test beam to End Station A. To keep this option alive, SLAC has made a proposal to add a pulsed extraction magnet, Be target and W collimator to the beamline. This would allow extraction in a variety of methods:
- The LCLS beam halo would impact on the W collimator, giving very low rate (~1 Hz) parasitic running of 2-6 GeV secondaries into the ESA line.
- Kick primary LCLS beam into the A line, giving $10^{10}$ Hz level beam.
- Kick primary LCLS beam into the Be target, allowing for pulsed secondaries.

End Station A is a very large area which can support many simultaneous experiments and will provide excellent user accessibility.

Fermilab: Fermilab's Meson Test Beam Facility remains in operation whenever the 120 GeV Main Injector is active. There is typically a 10 week shutdown period during the year. Beam is available in a 4 second flattop every minute, for 12 hours a day. Primary 120 GeV protons can be delivered at a rate of 200 KHz, or secondaries with any energy below 64 GeV can be tuned for. The rate of secondaries is a monotonically decreasing function of energy, with rates at the lowest energy of 1 GeV of around 4 KHz. The beam composition becomes dominated by electrons at around 8 GeV and lower. Fermilab has been supporting the CALICE calorimetry effort and has developed a variety of beam tunes for that collaboration. One notable achievement by the Accelerator Division this last year was the creation of a pulsed extraction mode, geared towards simulating the ILC beam structure. Figure 1 shows two pulses of 1 millisecond long beam, separated by 200 milliseconds. This mode is now available for users.

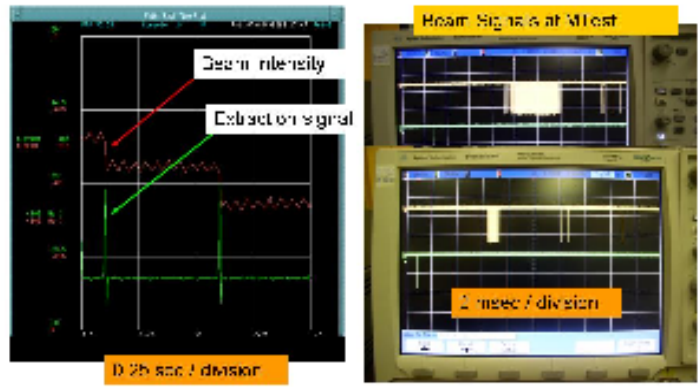

Figure 1. Pulsed extraction at Fermilab's test beam. Green line shows pulsed magnet current. Red line shows circulating beam. The oscilloscope shows beam scintillator pulses at the user facility for 1 msec and 5 msec extraction.

## 3   Future of Test Beam Facilities

### 3.1   Requirements for Tracking and Vertex Detectors (presentation by Ingrid Gregor of DESY)[4]

The requirements for tracking detectors at the ILC are significantly more stringent than what has been achieved to date: on the order of 5 μm for the impact parameter in the vertex detector (with a 0.1% radiation length per layer), and a momentum resolution in the tracker of $5 \times 10^{-5}$ (GeV/c)$^{-1}$ (with a 3% limit on total radiation length before the calorimeter). A variety of silicon pixel vertex detector solutions are envisioned, and both a silicon and TPC outer tracker are being planned for. All of these detectors will require high energy beam for resolution tests and eventually operation in a magnetic field for resolution studies.

The availability of the EUDET supported solenoid and pixel tracker at DESY allows for a considerable amount of detailed testing of silicon and TPC trackers for ILC right now. The high energy beams at CERN and Fermilab will also help cover the immediate need for proof-of-principle for the various technologies. However, the most challenging aspects of the tracking detectors for the ILC will be in how they contribute to a detailed understanding of jet reconstruction, since Higgs physics analysis relies heavily on jets.

To test these detectors in a realistic environment will require realistic magnetic fields. This means fields in the 3-5 Tesla range. In that range, Monte Carlo indicates that pixel track point resolution will change by 50%. To test vertex detectors, the bore size of the magnet need not be excessive, because of the small size of the test detectors. To test future TPC and silicon trackers, however, will require larger bore magnets – on the order of a meter or more. No current facility has such a large bore, high field magnet. The M1 magnet in the H2 beamline at CERN is 3 Tesla, with a 0.8 meter bore. If it becomes available, it is possible that this magnet would be applicable. If not, extra funds would be needed to either develop a new magnet, or modify an existing one. The time scale at which this kind of magnet is needed is several years from now.

A facility that can be made now, and which could be timely, is a jet testing facility. One could simulate a jet type of environment by having a high energy beam (>50 GeV) passing through a pixel telescope, and impacting on a series of targets inside a moderate magnetic field. Vertex detectors could sit just downstream of the targets, while trackers and calorimeters would be situated further downstream, allowing for curvature in the tracks. Both CERN and FNAL could host such a facility on a short time scale and one would imagine that it would be dedicated to ILC detector testing.

### 3.2 Requirements for Calorimetry (presentation by Andy White of U. Texas – Arlington)[5]

The great progress shown in calorimetry testing at DESY, CERN and FNAL indicates that current test beam facilities span the required energy and particle i.d. range. Test calorimeters have seen beam ranging from 0.5 to 200 GeV, with beam types including electrons, muons, pions and protons. The current facilities will continue to play a large role in testing modest scale (<1 cubic meter) calorimeter options.

However, to truly test ideas such as particle flow calorimetry, or dual readout calorimetry, large prototypes will need to be constructed to fully absorb hadronic showers. The CALICE calorimeter project has shown that a sizeable collaboration is needed to build such a device, transport it to facilities for long term testing and then analyzing the substantial amount of data produced. The next stage of calorimetry testing will be even larger, with realistic support structures involved, spanning several meters. A permanent home for these detectors will likely need to be found, with an ILC dedicated test beam line at CERN or Fermilab.

### 3.3 Putting it All Together

Ideally, in several years, the next generation of calorimeters will find their home in the same location as the 'jet' tracking detector test facility described above. In this fashion, integrated 'slice-tests' can be performed for a wide range of detectors and with a wide range of beam types. There is a proposal, called EUVIF, to integrate silicon detectors and a TPC into a unified framework, including the EUDET pixel telescope, and a common data acquisition and slow controls. Adding room for a calorimeter to this proposal, and making sure a significant magnetic field is included, seems like the path to take.